\documentclass[aps,pra,showpacs,groupedaddress,amsfonts,amssymb,amsmath]{revtex4}

\begin{document}

\title{A perturbative approach to a class of Fokker-Planck equations}

\author{Choon-Lin Ho}
\author{Yan-Min Dai}
 \affiliation{Department of Physics, Tamkang University,
 Tamsui 251, Taiwan, Republic of China}

\date{May 18, 2007}

\begin{abstract}
In this paper we present a direct perturbative method to solving
certain Fokker-Planck equations, which have constant diffusion
coefficients and some small parameters in the drift coefficients.
The method makes use of the connection between the Fokker-Planck
and Schr\"odinger equations.  Two examples are used to illustrate
the method. In the first example the drift coefficient depends
only on time but not on space.  In the second example we consider
the Uhlenbeck-Ornstein process with a small drift coefficient.
These examples show that the such perturbative approach can be a
useful tool to obtain approximate solutions of Fokker-Planck
equations with constant diffusion coefficients.

\end{abstract}


 \maketitle

\section{Introduction}

The Fokker-Planck (FP) equation is one of the basic tools used to
deal with fluctuations in various kinds of systems \cite{FP}.  It
has found applications in such diverse areas as physics,
astrophysics, chemistry, biology, finance, etc.  Owing to its wide
applicability, various methods of finding exact and approximate
solutions of the FP equations have been developed.  Most of the
methods, however, are concerned with FP equations with
time-independent diffusion and drift coefficients \cite{FP}.
Generally, it is not easy to find solutions of FP equations with
time-dependent diffusion and drift coefficients.

One of the methods of solving FP equation with time-independent
diffusion and drift coefficients is to transform the FP equation
into a Schr\"odinger-like equation, and then solve the eigenvalue
problem of the latter. The transformation to the Schr\"odinger
equation of a FP equation eliminates the first order spatial
derivative in the FP operator and creates a Hermitian spatial
differential operator.  This method is useful when the associated
Schr\"odinger equation is exactly solvable; for example with
infinite square well, harmonic oscillator potentials, etc. Several
FP equations have been exactly solved in this way \cite{FP}. The
method can also be useful for approximate results if a finite part
of the spectrum of the  associated Schr\"odinger equation can be
exactly known \cite{HS1}. When this happens the Schr\"odinger
equation, and hence the associated FP equation, are called
quasi-exactly solvable \cite{QES}.

In this paper we would like to show that the connection between
the FP and Schr\"odinger equations can also be useful even when
the drift coefficients are time-dependent.  Based on this
connection, we present a direct perturbative method which can be
used to solve a certain class of FP equations.

\medskip

\section{Perturbative approach to FP equations}

In one dimension, the FP equation of the probability density
$P(x,t)$ is \cite{FP}
\begin{gather}
\frac{\partial}{\partial t} W(x,t)=\left(-\frac{\partial}{\partial
x} D^{(1)}(x,t) + \frac{\partial^2}{\partial
x^2}D^{(2)}(x,t)\right)W(x,t). \label{FPE}
\end{gather}
The functions $D^{(1)}(x,t)$ and $D^{(2)}(x,t)$ in the FP equation
are, respectively,  the drift and the diffusion coefficient. The
drift coefficient represents the external force acting on the
particle, while the diffusion coefficient accounts for the effect
of fluctuation.  The drift coefficient is usually expressed in
terms of a drift potential $U(x,t)$ according to
$D^{(1)}(x,t)=-{\partial U(x,t)}/{\partial x}$.  In this paper we
shall be concerned with an important class of FP equations,
namely, those with constant diffusion coefficients
$D^{(2)}(x,t)=D>0$.

The FP equation is closely related to the Schr\"odinger equation.
A FP equation with time-independent drift
$D^{(1)}(x,t)=D^{(1)}(x)$ can be transformed via a similarity
transformation to a Schr\"odinger-like equation with
time-independent potential \cite{FP,HS1,HS2}.  Hence the FP
equation can be exactly solved if the associated Schr\"odinger
equation can be exactly solved.  It is also due to this connection
that time-independent WKB method can be applied to obtain
approximate solutions of the FP equation \cite{Caroli}.

When the drift coefficient of the FP equation is time-dependent,
the FP equation can still be transformed into a time-dependent
Schr\"odinger-like equation, only now the potential in its
associated Schr\"odinger equation is time-dependent.  We show
their connection below.

We first define $\psi(x,t)\equiv e^{U(x,t)/2D}W(x,t)$.
Substituting this into the FP equation, we find that $\psi$
satisfies the Schr\"odinger-like equation:
\begin{equation}
\frac{\partial\psi}{\partial t}= D\frac{\partial^2 \psi}{\partial
x^2}+\left(\frac{U^{\prime\prime}}{2} -\frac{U^{\prime 2}}{4D} +
\frac{\dot{U}}{2D}\right)\psi, \label{S-like}
\end{equation}
 where the prime and dot
denote the derivatives with respect to $x$ and $t$, respectively.
Eq.~(\ref{S-like}) is the time-dependent Schr\"odinger-like
equation associated with the FP equation.  Owing to the $\dot{U}$
term, the potential in the Schr\"odinger equation is
time-dependent.

If (\ref{S-like}) can be exactly solved, then the original FP
equation is also exactly solved. However, solving the
Schr\"odinger equations with time-dependent potentials is
generally difficult.  In most cases one has to resort to
approximate or numerical methods.

In this paper, we would like to show that a direct perturbative
approach can be useful to solve the FP equation containing a small
parameter in the drift potential.

We  shall consider drift potential of the form
$U(x,t)=\sum_{n=0}^\infty \lambda^n U_n(x,t)$, where
$|\lambda|<<1$ is a small parameter. We introduce the ansatz
$\psi=\exp(S(x,t,\lambda)/D)$, where $S$ is expanded in powers of
$\lambda$:
\begin{eqnarray}
S(x,t,\lambda)=\sum_{n=0}^\infty \lambda^n S_n(x,t).\label{S-exp}
\end{eqnarray}
From (\ref{S-like}) we find that $S(x,t,f)$ satisfies
\begin{eqnarray}
\dot{S}&=&DS^{\prime\prime}+ S^{\prime 2} + \overline{U},
\label{S-eq}\\
\overline{U}&\equiv & \frac{D}{2}U^{\prime\prime}
-\frac{1}{4}U^{\prime 2} + \frac{1}{2}\dot{U}.\label{Ubar}
\end{eqnarray}
Substituting  (\ref{S-exp}) into (\ref{S-eq}) and collecting terms
of the same order in $\lambda$, we arrive at a set of differential
equations determining the functions $S_n$:
\begin{gather}
\dot{S}_0 =DS_0^{\prime\prime} + S_0^{\prime
2}+\frac{D}{2}U_0^{\prime\prime}
-\frac{1}{4}U_0^{\prime 2} + \frac{1}{2}\dot{U}_0,\label{S0}\\
\dot{S}_1 =DS_1^{\prime\prime} + 2 S_0^\prime S_1^\prime + \mbox{
terms in $\overline{U}$ of the order $\lambda$},\label{S1}\\
\dot{S}_2 = DS_2^{\prime\prime} + 2 S_0^\prime S_2^\prime +
S_1^{\prime 2} + \mbox{
terms in $\overline{U}$ of the order $\lambda^2$},\label{S2}\\
\vdots \nonumber\\
\dot{S}_n=DS_n^{\prime\prime} + \sum_{k=0}^n S_k^\prime
S_{n-k}^\prime  + \mbox{ terms in $\overline{U}$ of the order
$\lambda^n$},~~~n\geq 0.\label{Sn}
\end{gather}
We note that the probability density $W_0=\exp(S_0/D)$ with $S_0$
satisfying (\ref{S0}) is the solution of the FP equation with
constant $D$ and drift potential $U_0(x,t)$.

By solving $S_0$, $S_1$, $S_2,\ldots$  from the above equations,
we can obtain an approximate solution of $\psi$, and hence of
$W(x,t)$.  We shall illustrate this by two simple examples below.

\section{FP equations with drift potential $U(x,t)=\lambda xV(t)$}

Let us consider a class of FP equation with a weak drift potential
of the form $U(x,t)=\lambda xV(t)$, where $V(t)$ is some finite
function of time and $|\lambda|<<1$. In this case,
$U_0=U_n=0~(n\geq 2)$. This FP equation can thus be viewed as a
diffusion equation (or Wiener process) perturbed by a weak
time-dependent drift potential.

Let us assume the initial profile of $W$ to be the delta-function,
i.e. $W(x,t)\to \delta(x)$ at $t=0$ as $\lambda \to 0$. Then the
solution of (\ref{S0}) is
\begin{eqnarray}
S_0(x,t)=-\frac{D}{2}\ln(4\pi Dt) - \frac{x^2}{4t},\label{S-0}
\end{eqnarray}
giving the probability density
\begin{eqnarray}
W_0=e^{\frac{S_0}{D}}=\frac{1}{\sqrt{4\pi
Dt}}\exp\left(-\frac{x^2}{4Dt}\right). \label{W0}
\end{eqnarray}
We have $\int_{-\infty}^{\infty} W_0 dx=1$. This is the well known
solution of the diffusion equation having the delta-function as
its initial profile.

$S_1(x,t)$ is determined from $S_0$ by (\ref{S1})
\begin{eqnarray}
\dot{S}_1=DS^{\prime\prime}_1 -\frac{x}{t} S^\prime_1 +
\frac{x}{2}\dot{V}.\label{S-1}
\end{eqnarray}
It is easily checked that (\ref{S-1}) is solved by
\begin{eqnarray}
S_1(x,t)&=&\frac{x}{2} V(t) -\frac{x}{2t}\overline{V}(t) +
c_0\\
\overline{V}(t)&\equiv & \int^t V(t)dt + c_1.
\end{eqnarray}
Here $c_0$ and $c_1$ are arbitrary real constants.  $c_0$ can be
absorbed into the normalization constant of $W(x,t)$. We demand
that $c_1$ be so chosen  as to render $S_1$ finite in the limit
$t\to 0$.  For instance,  if $V(t)=\cos(\omega t)$, we shall
choose $c_1=0$, whilst for $V(t)=\sin(\omega t)$ we have
$c_1=1/\omega$.

For $S_2(x,t)$, it is determined by (\ref{S2})
\begin{eqnarray}
\dot{S}_2=DS^{\prime\prime}_2 -\frac{x}{t} S^\prime_2 +
\frac{1}{4}\left(V(t) - \frac{1}{t}\overline{V}(t)\right)^2
-\frac{V(t)^2}{4}.\label{S-2}
\end{eqnarray}
Solution of (\ref{S-2}) is
\begin{eqnarray}
S_2(x,t)=-\frac{1}{4t}\overline{V}(t)^2 + c_3\frac{x}{t}+c_2.
\end{eqnarray}
Again, $c_2$ can be absorbed into the normalization constant. We
demand that $S_2$ be finite for all $x$ as $t\to 0$, hence we set
$c_3=0$. With these forms of $S_0,~S_1$ and $S_2$, one finds from
(\ref{Sn}) that $S_3$ ($n\geq 3$) satisfies the equation:
\begin{eqnarray}
\dot{S}_3=DS^{\prime\prime}_3 -\frac{x}{t} S^\prime_3,\label{S-3}
\end{eqnarray}
which is solved by $S_3=c_5 x/t+c_4$ with arbitrary constants
$c_4$ and $c_5$. Again finiteness of $S_3$ as $t\to 0$ rules out
the $x/t$ term, and the constant term can be absorbed into the
normalization constant, and hence we set $c_4=c_5=0$. Continuing
this process, we find that all subsequent $S_n$'s also satisfy
(\ref{S-3}), and hence all $S_n$ must be taken as constant, which
can all be set to zero.

Hence, in this example we can solve all the $S_n$ from the set of
equations (\ref{S1}), (\ref{S2}) to (\ref{Sn}).  The probability
density $W(x,t)$ is given by the following expression:
\begin{eqnarray}
W(x,t)&=&e^{-\frac{\lambda x}{2D}V(t)}e^{(S_0+\lambda S_1+
\lambda^2 S_2)/D}\nonumber\\
&=&\frac{1}{\sqrt{4\pi Dt}}\exp\left(-\frac{1}{4Dt}\left(x+\lambda
\overline{V}(t)\right)^2\right). \label{W}
\end{eqnarray}
$W$ in (\ref{W}) is normalized.  One notes that (\ref{W}) is very
similar to the solution (\ref{W0}) of the diffusion equation, only
with the $x$ in $W_0$ being replaced by $\bar{x}\equiv x+\lambda
\overline{V}(t)$. It turns out that $W$ in (\ref{W}) is indeed the
exact solution of the FP equation with $U(x,t)=\lambda  xV(t)$ for
any value of $\lambda$.  In fact, in terms of the new variable
$\bar{x}$ and $t$, the FP equation reduces to the diffusion
equation:
\begin{eqnarray}
\frac{\partial W(\bar{x}, t)}{\partial t} = D \frac{\partial^2
W(\bar{x}, t)}{\partial \bar{x}^2}.
\end{eqnarray}
Eq.~(\ref{W}) is the exact solution of this diffusion equation
with initial profile $\delta (x+ \lambda \overline{V}(0))$.  The
peak of $W(x,t)$ shifts according to the function $\lambda
\overline{V}(t))$. For instance, if $V(x)\sim \sin(\omega t)$ or
$\cos(\omega t)$, the peak oscillates about $x=0$ as $W(x,t)$
spreads outward.

\section{Uhlenbeck-Ornstein process with a small drift potential}

Next we consider  the Uhlenbeck-Ornstein process \cite{UO,FP},
which is described by the FP equation with constant diffusion
coefficient and a time-independent drift potential $U(x)=\lambda
x^2/2$.  This process is exactly solvable, with the probability
density given by
\begin{eqnarray}
W(x,t)=\sqrt{\frac{\lambda }{2\pi D(1-e^{-2\lambda t})}}
\exp\left(-\frac{\lambda  x^2}{2 D(1-e^{-2\lambda
t})}\right).\label{W-UO}
\end{eqnarray}
We would like to apply the above perturbative approach to this
process when $\lambda$ is small.  Hence we are treating the
Uhlenbeck-Ornstein process as a perturbed Wiener process.

The associated Schr\"odinger equation (\ref{S-like}) and
$\overline{U}$ in (\ref{Ubar}) do not contain $\dot{U}$ in this
case. $S_0$ is still to be solved from (\ref{S0}).  If we assume
an initial profile $W(x,0)\to \delta (x)$ as $\lambda \to 0$, then
$S_0$ is again given by (\ref{S-0}).

With this $S_0$, (\ref{S1}) and (\ref{S2}) read
\begin{gather}
\dot{S}_1 =DS_1^{\prime\prime}  -\frac{x}{t} S_1^\prime +
\frac{D}{2},
\label{UO1}\\
\dot{S}_2 = DS_2^{\prime\prime} -\frac{x}{t} S_2^\prime +
S_1^{\prime 2} - \frac{x^2}{4}.\label{UO2}
\end{gather}
Eq.~(\ref{UO1}) is solved by
\begin{eqnarray}
S_1=\frac{1}{2}Dt + c_1\frac{x}{t} + c_0,~~\mbox{$c_0,~c_1$:
arbitrary constants}.
\end{eqnarray}
As before, finiteness of $S_1$ requires $c_1=0$, and $c_0$ is
absorbed into normalization constant. Plugging $S_1=Dt/2$ into
(\ref{UO2}), we obtain the solution for $S_2$:
\begin{eqnarray}
S_2=-\frac{1}{12}Dt^2 - \frac{1}{12}x^2t +
c_3\frac{x}{t}+c_2,~~\mbox{$c_2,~c_3$: arbitrary constants}.
\end{eqnarray}
By the same reasons given before, we set $c_3=0$ and leave out
$c_2$. Putting all the solutions together, we have, up to the
$\lambda^2$ terms, the approximate expression for $W$
\begin{eqnarray}
W(x,t)\approx \frac{1}{\sqrt{4\pi
Dt}}\exp\left(-\frac{x^2}{4Dt}\left(1+\lambda
t+\frac{1}{3}\lambda^2 t^2\right)+\frac{1}{2}\lambda
t-\frac{1}{12}\lambda^2 t^2\right).\label{W-UO1}
\end{eqnarray}
Now using the series expansion of $\ln(1+x)\approx x-x^2/2
+\ldots$ for small $x$, we can rewrite the last two terms in the
exponent of (\ref{W-UO1}) as
\begin{eqnarray}
\frac{1}{2}\lambda t-\frac{1}{12}\lambda^2 t^2\approx
\frac{1}{2}\ln\left(1+\lambda t +\frac{1}{3} \lambda^2 t^2\right).
\end{eqnarray}
Then (\ref{W-UO1}) becomes
\begin{eqnarray}
W(x,t)\approx \sqrt{\frac{1+\lambda t + \frac{1}{3}\lambda^2
t^2}{4\pi Dt}}\exp\left(-\frac{x^2}{4Dt}\left(1+\lambda
t+\frac{1}{3}\lambda^2 t^2\right)\right).\label{W-UO2}
\end{eqnarray}
$W$ given by (\ref{W-UO2}) is normalized.  It is easily checked
that (\ref{W-UO2}) is the same as the approximate expression
obtained from (\ref{W-UO}) when the factors in the square-root and
the exponent are expanded up to $\lambda^2$ terms.  Hence, the
result obtained by the perturbative approach is consistent with
the exact result.

\section{Summary}

In summary, we have applied a direct perturbative theory to solve
certain FP equations, which have constant diffusion coefficients.
Two examples are used to illustrate the method.  In the first
example the drift coefficient depends only on time but not on
space.  In the second example we treat the Uhlenbeck-Ornstein
process with a small drift coefficient. These examples demonstrate
that such perturbative approach is feasible for obtaining
approximate solutions of FP equations.

\newpage

\begin{acknowledgments}

This work is supported in part by the National Science Council
(NSC) of the Republic of China under Grant NSC 95-2112-M-032-012.

\end{acknowledgments}

\end{document}